\documentclass[dvips]{article}
\usepackage{icrctc07}

\title{The TeV Energy Spectrum of Mrk 421 Measured in A High Flaring State}
\shorttitle{TeV Energy Spectrum of Mrk 421}
\authors{A. Konopelko$^{1}$, W. Cui$^{1}$, C. Duke$^{2}$, J.P Finley$^{1}$}
\shortauthors{A. Konopelko and et al}
\afiliations{$^1$Purdue University, Department of Physics, 525 Northwestern Avenue, 
West Lafayette, IN 47907-2036, $^2$Grinnell College, Department of Physics, 
1116 8th Ave, Grinnell, IA 50112-1690}
\email{akonopel@purdue.edu}

\abstract{The BL Lac object (blazar) Mrk 421 was observed during its outburst in 
April 2004 with the Whipple 10 m telescope for a total of about 24.5~hours. The 
measured $\gamma$-ray rate varied substantially over the range from 4 to 
10~$\gamma$'s/min and eventually exceeded the steady $\gamma$-ray rate of 
the Crab Nebula (standard candle) by a factor of 3. The overall significance of the 
$\gamma$-ray signal exceeded 70~$\sigma$ and the total number of excess events 
was more than 10,000. The signal light curve does not show any particular 
variability pattern. This unique Mrk 421 outburst enabled the measurement of a high quality 
spectrum of very high-energy $\gamma$ rays in a high state of emission. This 
spectrum is a power-law and it extends beyond 10 TeV.}

\begin{document}
\maketitle

\section{Introduction}
\vspace*{-2mm}

Mkn~421 is the first detected, the closest known (red shift z = 0.030), and one of the best 
studied TeV $\gamma$-ray emitting blazars. The very high-energy (VHE) $\gamma$-ray 
emission arises from the particles accelerated in a relativistic jet directed along our line-of-sight. 
Since its discovery, Mkn~421 has shown very low baseline TeV $\gamma$-ray emission with 
a few extremely rapid flares on time-scales from one day to 15~minutes \cite{ref6}. 
In 2000 and 2001 Mkn~421 went into a flare state with an average flux of 4 times that of the 
Crab Nebula. Data taken during this flare have been used to extract the energy spectrum at 
high energies, up to 20 TeV. Noticeable variations in the hardness of the TeV $\gamma$-ray 
energy spectrum have been reported \cite{ref1,ref9}. A number of successful multi-wavelength 
campaigns for Mkn 421 have revealed evidence of a correlation of the simultaneously measured fluxes 
in X-ray and TeV $\gamma$-ray energy bands \cite{ref81,ref3}.

A one-zone synchrotron-self-Compton (SSC) model described the observational results 
reasonably well (e.g., \cite{ref8}). During April 2004, an intensive multi-wavelength 
monitoring campaign on the TeV blazar Mkn~421 has been performed simultaneously in radio, 
optical, X-ray, and $\gamma$ rays. The source was seen to be active in X-rays and TeV $\gamma$ 
rays with the peak flux exceeding the 3 Crab level \cite{ref3}. The time-averaged 
energy spectrum of Mkn~421 during the flaring state has been measured at high energies 
with H.E.S.S. using large zenith angles observations \cite{ref2}, and MAGIC \cite{ref22}. The Whipple 
10~m telescope was observing Mkn~421 extensively during the April 2004 flare. Here we report the 
results on the TeV $\gamma$-ray energy spectrum of Mkn~421 measured in a high flaring state 
during the April 2004 outburst with the Whipple 10~m telescope.

\vspace*{-3mm}
\section{The 10 m Whipple Telescope}
\vspace*{-2mm}

\begin{figure}[!t]
\includegraphics [width=0.48\textwidth]{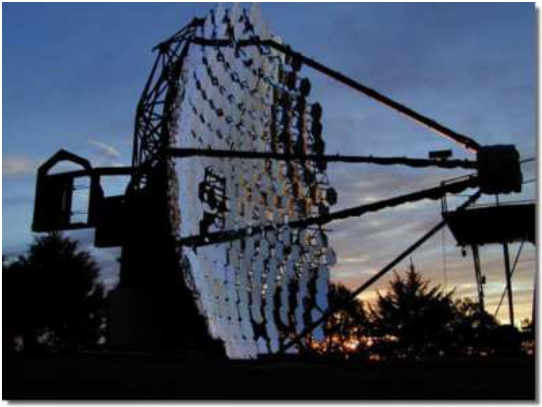}
\caption{Whipple Observatory 10 m Telescope at Mount Hopkins, Arizona.}
\label{fig1}
\vspace{-3mm}
\end{figure}

The reflector of the Whipple Observatory imaging atmospheric Cherenkov telescope (IACT)
(see Figure~\ref{fig1}) is a tessellated structure consisting of 248 spherical mirrors, which are 
hexagonal in shape and 61~cm from apex to apex, arranged in a hexagonal pattern. The 
mirrors are mounted on a steel support structure, which has a 7.3~m radius of curvature with 
a 10~m aperture. Each individual mirror has $\sim$14.6~m radius of curvature and is pointed 
toward a position along the optical axis at 14.6~m from the reflector. This arrangement 
constitutes a Davies, Cotton design \cite{ref4} of the optical reflector. The point-spread function 
of the telescope has a FWHM of $\sim 7.2'$ on-axis. 

In 1999, a high-resolution camera (GRANITE III) was installed at the telescope \cite{ref5}. 
It consists of 379 photo-multiplier tubes (PMTs) in a close packed hexagonal arrangement 
(each PMT subtending 0.11$^\circ$ on the sky) and has a 2.6$^\circ$ diameter. A set of light 
concentrators is mounted in front of the pixels to increase the light-collection efficiency by 
$\sim$38\%. The camera triggers if the signal in each of at least 3 neighboring PMTs out of 
the inner 331 exceeds a threshold of 32 mV, corresponding to $\sim$8-10 photo-electrons. 
The post-GRANITE III upgrade trigger rate of the Whipple Observatory 10~m telescope is 
$\sim$20-30 Hz at zenith.

\vspace*{-3mm}
\section{Observations \& Analysis}
\vspace*{-2mm}

Mkn~421 was observed with the Whipple Observatory 10~m IACT for about 24.5 hours 
of on-source data between April 9th and April 23d, 2004. Data have been taken in good 
weather at zenith angles less than 30$^\circ$ (mean elevation is about 75$^\circ$). 
During observations the raw detection rate was about 22~Hz. Observations have been 
performed using tracking mode in data runs of 28~minutes each. The recorded images 
are first flat-fielded using nightly measured nitrogen arc lamp PMT responses and then 
cleaned by applying a standard picture and boundary technique with canonical thresholds 
of 4.25 and 2.25 times the standard deviation of the PMT pedestal distributions, respectively 
(see, e.g., \cite{ref10}). To characterize the shape and orientation of calibrated images, the 
standard second-moment image parameters are calculated as described in \cite{ref14}. 
The {\it CAnalyze} package developed at Purdue University \cite{ref11} is used for the primary 
data analysis. Analysis methods known as Supercuts (see Table~\ref{t1}) and 
extended Supercuts are applied for the $\gamma$-ray/hadron separation. The latter utilizes 
dynamic, energy dependent orientation and shape cuts. The significance map 
of the sky region around Mkn~421 for the data set corresponding to a high emission state 
(exposure of 55~minutes) is shown in Figure~\ref{fig3}. The energy spectrum of Mkn~421 
is reconstructed using algorithms described in \cite{ref13}.    

\begin{figure}[!t]
\includegraphics [width=0.44\textwidth]{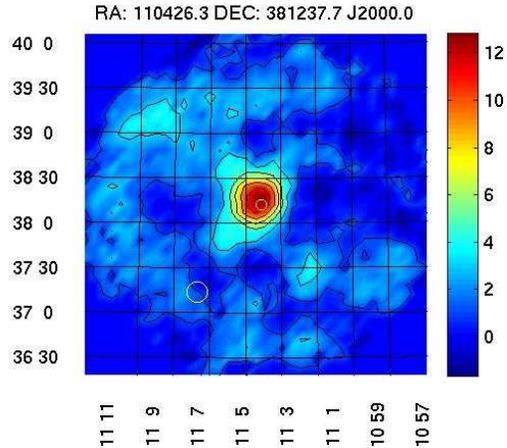}
\vspace*{-2mm}
\caption{The two-dimensional map of excess significances over the sky 
region around Mkn~421 generated for the data sample corresponding to 
a high emission state. The map of uncorrelated rectangular angular bins 
of a $0.1^\circ$ size has been smoothed using Gaussian kernel with the 
angular size of 0.22$^\circ$. The maximal significance of $\gamma$-ray 
excess using this analysis is 12.3~$\sigma$. }
\label{fig2}
\vspace*{-3mm}
\end{figure}

\begin{table}[!t]
\centering
\caption{{\it Supercuts} selection criteria for the Whipple Observatory 10~m telescope ({\it dc} 
stands for digital counts).}
\vspace*{2mm}
\begin{tabular}{lc}\hline
Quantity   & Image Parameter Cut \\ \hline 
Trigger & 1st \& 2nd brightest pixel $>$ 30 dc \\
Shape & $0.05^\circ < width < 0.12^\circ$ \\
             & $0.13^\circ < length < 0.25^\circ$ \\
Muon cut & $length/size < 0.0004 \, \rm (^\circ\,dc^{-1})$ \\
Quality cut & $0.4^\circ < distance < 1.0^\circ$ \\
Orientation & $\alpha < 15^\circ$ \\ \hline
\end{tabular}
\label{t1}
\vspace{-3mm}
\end{table}

\vspace*{-3mm}
\section{Simulations}
\vspace*{-2mm}

The KASCADE shower simulation code \cite{ref7} is used for generating 
the $\gamma$-ray and cosmic-ray induced air showers within the corresponding 
range of zenith angles and in the primary energy range between 50 GeV and 
100 TeV, assuming the $\gamma$-ray energy spectrum to be a power-law of 2.5. 
Simulations of the response of the Whipple Observatory 10~m telescope are 
carried out using the {\it GrISU} code, developed by the Grinnell College and Iowa 
State University groups.

\begin{figure}[!t]
\includegraphics [width=0.46\textwidth]{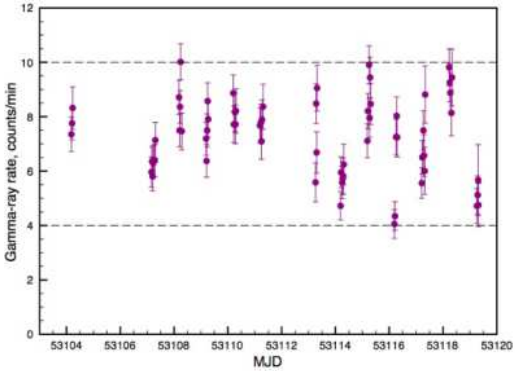}
\caption{Light curve of TeV $\gamma$-ray emission from Mkn~421 during its 
outburst in April 2004.}
\label{fig3}
\vspace{-3mm}
\end{figure}

\begin{table*}[!t]
\centering
\caption{Summary of Mkn~421 data analyzed by standard supercuts criteria (see Table~\ref{t1}.) }
\vspace*{2mm}
\begin{tabular}{rcrrrcc}\hline
                 & Time [min] & ON & OFF & Excess & $R_{\gamma}\, \rm min^{-1}$ & S/N [$\sigma$] \\ \hline 
April 2004 & 1472 & 18383 & 6872 & 11511 & 7.82 & 71.82 \\ 
High state & 55 & 825 & 280 & 545 & 9.85 & 16.40 \\  \hline
\end{tabular}
\label{t2}
\vspace{-2mm}
\end{table*}

\vspace*{-3mm}
\section{Results}
\vspace*{-2mm}

\begin{figure}[!t]
\includegraphics [width=0.43\textwidth]{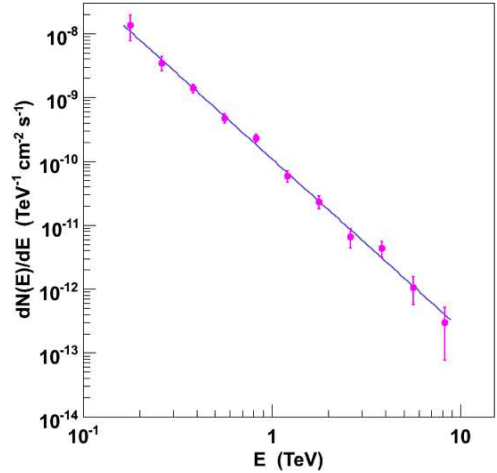}
\caption{The energy spectrum of Mkn~421 measured in 
a  high emission state during the outburst in April 2004.}
\label{fig4}
\vspace{-3mm}
\end{figure}

The light curve of Mkn~421, measured with the Whipple Observatory 10~m 
telescope during the outburst in April 2004, does not reveal any particular 
pattern (see Figure~\ref{fig2}). Two data runs with a flux above 3 Crab and 
an exposure of 55~minutes give exceptionally high $\gamma$-ray rate with 
an average value of 9.85 min$^{-1}$ and a total of 545 excess counts. This 
excess corresponds to a significance of the $\gamma$-ray signal at a level of 
16.4~$\sigma$. The corresponding $\gamma$-ray energy spectrum is consistent 
with a pure power-law of index 2.66. This spectrum does not indicate any apparent 
gradual change of slope (see Figure~\ref{fig4}). Such unusual behavior of 
the Mkn~421 TeV $\gamma$-ray spectrum measured in a high emission state 
suggests that the previously observed curvature in Mkn~421 $\gamma$-ray 
spectrum is an intrinsic feature of the source rather than the footprint of 
$\gamma$-ray absorption by interacting with the soft photons of extragalactic 
background light.

\vspace*{-3mm}
\section{Modeling}
\vspace*{-2mm}

\begin{figure}[!t]
\includegraphics [width=0.44\textwidth]{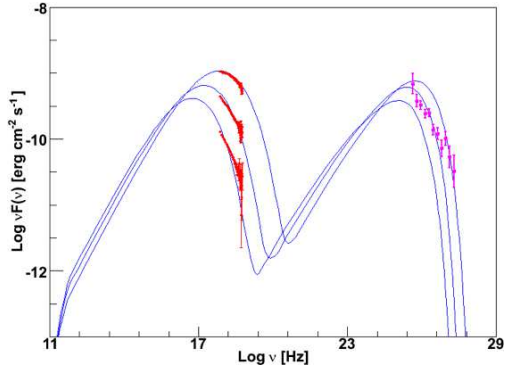}
\caption{The spectral energy distribution of the Mkn~421 emission. The X-ray data, 
corresponding to low, medium, and high emission states have been taken 
from \cite{ref3}.}
\label{fig5}
\vspace{-3mm}
\end{figure}

To model the multi-wavelength spectrum of the BL Lac object Mkn~421 
in a homogeneous Synchrotron Self Compton (SSC) scenario 
we use an approach described in \cite{ref12}. This 
method involves prescribing an injection function for relativistic electrons 
and solving the two time-dependent kinetic equations for the electron 
and photon distributions of the source. All relevant physical processes are taken 
into account in the code, i.e., synchrotron radiation, inverse Compton 
scattering, photon-photon pair production, and synchrotron self-absorption. 
Seven model parameters are required to specify a source in a stationary state. 
These are the Doppler factor $\delta$, the radius $R$ of the source, the 
magnetic field strength $B$, the index {\it s} of the electron injection spectrum, 
the Lorentz factor at the cut-off of the injection spectrum $\gamma_{max}$, 
the amplitude of the injection spectrum $q_e$, and the effective escape time 
of relativistic electrons $t_{esc}$. To optimize a fit to a particular data set, 
we used physically motivated starting values for these seven parameters
(see \cite{ref8}). This is done by identifying six scalars characterizing the 
blazar spectrum, the peak frequency $\nu_{sync}$
of the synchrotron emission, the peak frequency $\nu_{compt}$ of the inverse 
Compton emission, the total nonthermal luminosity $L$, the approximate 
ratio $\eta$ of the total flux in the inverse Compton part to that in the 
synchrotron part of the spectrum, the break frequency $\nu_{break}$,
the low-frequency spectral index, and finally the fastest variability timescale $t_{var}$.
These observables enable one to find reasonable starting values of the
seven parameters of the SSC model. Using this approach we tried to fit the 
measured spectral energy distribution  for a high state of Mkn~421. 
Despite that the X-ray spectra can be 
reasonably well reproduced by the one-zone SSC model, it does not provide a 
satisfactory fit to the TeV $\gamma$-ray spectrum of Mkn~421 measured in 
a high emission state (see Figure~\ref{fig5}). It is worth noting that unfolding 
the TeV $\gamma$-ray spectrum from the IR absorption (see \cite{ref8} for 
further details) yields an energy spectrum substantially harder that the 
measured one, which is more difficult to interpret in the framework of the one-zone 
SSC model.

\vspace*{-3mm}
\section{Discussion}
\vspace*{-2mm}

Observations of Mkn~421 with Whipple \cite{ref9} and HEGRA \cite{ref1} 
during its historical flare in 2000 and 2001 have been used to extract the 
energy spectrum at high energies, up to 20~TeV. As stated by both groups, 
the energy spectrum of Mkn~421 is evidently curved. Analysis of the data 
revealed significant variations of the spectral slope at energies below 3~TeV 
depending on the emission state. However the best empirical fit to the data 
taken in both the high and the low states gives the same cut-off energy of
about 3.6~TeV. Recently, the gradual steepening in the Mkn~421 TeV energy 
spectrum was evidently detected with HESS \cite{ref2} and Magic \cite{ref22} 
during their 2004 and 2005 observational campaigns. This is consistent with 
the assumption that all Mkn~421 spectra are affected by intergalactic absorption. 
Here we report on the TeV $\gamma$-ray spectrum as measured with the 
Whipple 10~m telescope during a high emission state (at the flux level of 3 Crab), 
which does not indicate a cut-off feature in muti-TeV energy range and can 
be well-fitted by a pure power-law. Such an unusual spectrum indicates that the 
curvature could be an intrinsic feature of the source. This spectrum offers a 
challenge for the one-zone SSC model, and it may severely constrain the 
spectral energy distribution of extragalactic background light, which is the 
subject of a forthcoming refereed publication.



\vspace*{-3mm}

\bibliographystyle{plain}

\end{document}